\documentclass[conference]{IEEEtran}
\normalsize
\usepackage{color}
\usepackage{bbm}
\usepackage{gensymb}
\usepackage{extarrows}
\usepackage{url}
\usepackage{balance}
\usepackage{graphicx} 
\usepackage{epstopdf}
\usepackage{stfloats} 

\usepackage{amsfonts}
\usepackage{cite}
\usepackage{graphicx}
\usepackage{amsmath}
\usepackage{times}
\usepackage{latexsym}
\usepackage{subfigure}
\usepackage{bm}
\usepackage{amssymb}
\usepackage{cases}
\usepackage{array}
\usepackage{flushend}
\usepackage{fancyhdr}
\usepackage{stfloats}
\usepackage{setspace}
\usepackage{flushend}
\usepackage{mathrsfs}
\usepackage{amsthm} 
\usepackage{multirow}
\usepackage{lipsum}
\usepackage{extarrows}
\usepackage{epsfig}
\usepackage{multicol}
\usepackage{subfigure}
\usepackage[square, comma, sort&compress, numbers]{natbib}
\usepackage{mathtools}
\DeclarePairedDelimiter\ceil{\lceil}{\rceil}

\graphicspath{{figure/}}
\title{MmWave V2X Beam Training with Situation Awareness}
\begin{document}
\title {MmWave Beam Prediction with Situational Awareness: A Machine Learning Approach}
	\author{\IEEEauthorblockN{Yuyang Wang$\stackrel{\ddag}{,}$ Murali Narasimha$\stackrel{\S}{,}$  and Robert W. Heath Jr.$\stackrel{\ddag}{}$}
		\IEEEauthorblockA{$^\ddag$ Department of Electrical and Computer Engineering, The University of Texas at Austin, USA \\
			$^\S$Huawei Technologies, USA\\
			Email: $\{$yuywang, rheath$\}$@utexas.edu, murali.narasimha@huawei.com }}
	\IEEEspecialpapernotice{(Invited Paper)}
\maketitle 

\begin{abstract} 
Millimeter-wave communication is a challenge in the highly mobile vehicular context. Traditional beam training is inadequate in satisfying low overheads and latency. In this paper, we propose to combine machine learning tools and \emph{situational awareness} to learn the beam information (power, optimal beam index, etc) from past observations. We consider forms of situational awareness that are specific to the vehicular setting including the locations of the receiver and the surrounding vehicles. We leverage regression models to predict the received power with different beam power quantizations. The result shows that situational awareness can largely improve the prediction accuracy and the model can achieve throughput with little performance loss with almost zero overhead.  \end{abstract} 

\section{Introduction} 
Configuring millimeter wave (mmWave) antenna arrays is a challenging task in vehicular applications \cite{VaChoShi:Inverse-Multipath-Fingerprinting:17}, \cite{Wan:Beam-codebook-based:09}. With high mobility, vehicles suffer from intermittent blockages from trucks, buses, etc, and therefore potential coverage holes, which requires frequent beam re-alignment to maintain transmission links with high data rates \cite{WanVenMol:Blockage-and-Coverage-Analysis:17}. Current solutions adopted in IEEE 802.11ad is inadequate in satisfying the low overheads and latency requirement in many vehicular applications \cite{SurVenZha:60-GHz-indoor-networking-through:15}, \cite{NitCorFlo:IEEE-802.11-ad:-directional:14}. MmWave, though, is the only viable solution to support massive data sharing and is incredibly valuable for infotainment services, with the emerging applications in 5G vehicular/cellular communication \cite{ChoVaGon:Millimeter-Wave-Vehicular-Communication:16,RapHeaDan:Millimeter-wave-wireless:14,RapSunMay:Millimeter-wave-mobile:13}. Consequently, low overhead methods with high efficiency and robustness need to be designed to enable fast configuration of millimeter wave links. 

One alternative to simplify beamforming in mmWave vehicular networks is to make use of out-of-band side information \cite{PreAliVa:Millimeter-Wave-communication:17}.  \emph{Side information} is ubiquitous in intelligent transportation systems. Vehicles have a number of sensors including GPS, radar, LIDAR, and cameras. In addition, connectivity between vehicles allows exchange of such information, which equips the vehicles with \emph{situational awareness} \cite{WeiSniKim:Towards-a-viable-autonomous:13}, \cite{ RasGre:Automotive-radar-and-lidar:05}. In \cite{PreAliVa:Millimeter-Wave-communication:17}, \cite{nitsche2015steering},  out-of-band  information assisted mmWave  beam training was proposed to leverage the data from sensors or 
other communication systems.  In \cite{AliPreJr.:Millimeter-Wave-Beam-Selection:18}, it was argued that there exists congruency between the channel at mmWave and sub-6 GHz bands, which can be leveraged to do beam selection and channel estimation. In \cite{GonMenHea:Radar-aided-beam:16}, a beam alignment solution was designed by extracting useful information from radar signal to configure the antennas and design beams at vehicles. An inverse fingerprinting approach was proposed to facilitate optimal beam pair selection in \cite{VaChoShi:Inverse-Multipath-Fingerprinting:17}. Given the receiver location, the infrastructure recommends and ranks the beams based on the occurrences of optimal beam pair in the dataset for that location. In \cite{KlaBatPre:Ray-Tracing-5G-MIMO-Data:16}, a framework of generating 5G MIMO dataset using ray tracing was proposed, and deep learning model was applied to assist in mmWave beam selection with temporally-correlated vehicle moving trajectories.

In this paper, we propose a novel framework to leverage machine learning tools with the availability of \emph{situational awareness}, in predicting mmWave beam power. In the vehicular contexts, road side buildings and infrastructures are stationary, and pedestrians are small in size, which makes vehicles the most important mobile reflectors in the urban canyons. The situational awareness of the vehicles, therefore, can be mapped to the received power of different beams. We propose to use the vehicle locations as features to predict the received power of \emph{any beam} in the beam codebook, with low or almost-zero feedback overhead. Vehicle locations may be obtained from the basic safety message in dedicated short-range communications (DSRC), or through similar functionality in a cellular system.  First, we apply different regression models over the \emph{strongest beam power} \cite{url_dataset}. We compare the results with different levels of situational awareness and show that full situational awareness can largely improve the prediction accuracy. We also investigate how different channel quality indicator (CQI) quantization parameters can effect the results using our specific dataset. We show that the optimal parameters depend on the dataset statistics and CQI quantization does not degrade the performance much, if high resolution can be guaranteed. Lastly, we evaluate the performance of multi-variate regression over power of all beam pairs. We observe that beam selection based on the power prediction can achieve higher throughput compared to that based on classification.

\section{Database establishment}
\subsection{Simulation setup}
\begin{figure}[h]
	\centering
	\includegraphics[width = 1.8in]{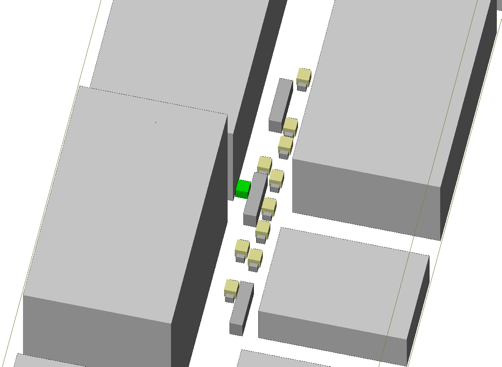}
	\caption{An illustration of the urban canyon in the ray tracing simulation. Fixed buildings are deployed at the roadside. The green box represents the RSU, and the receivers (yellow boxes) are mounted on top of the low-height cars. Vehicles are randomly dropped on the two lanes, under some certain vehicle density and truck-to-car ratio. }\label{fig:illustration}
\end{figure}

In this paper, we set the analysis in a two-lane straight street in the urban canyon, as shown in Fig. \ref{fig:illustration}. Building placements are predetermined. The road-side unit (RSU) is deployed at the road side, at the height of $5$ meters, and there are two types of vehicles in the environments, respectively the trucks (with identical sizes of length, height, width = $T_\ell, T_h, T_w$) and low height cars with size of $C_\ell, C_h, C_w$. We use \emph{Wireless Insite} from \emph{Remcom} to obtain the channel and beam information  \cite{url_remcom}. We only simulate the channels of low-height cars since high trucks are free of blockage, and the optimal beam pair is always line-of-sight (LOS).
 \subsection{Channel model}\label{sec:beamtraining} 
We generate the mmWave channel by combining the outputs from ray tracing and the geometric channel model \cite{VaChoShi:Inverse-Multipath-Fingerprinting:17}. The \emph{rays} in ray tracing are equivalent to the \emph{paths} in mmWave channel modeling. From the ray tracing output, we obtain the path information $\{\phi_\ell^A, \theta_\ell^A, \phi_\ell^D, \theta_\ell^D, \tau_\ell, a_\ell\}, ~\ell\in\{1, 2\cdots, L\}$ of the strongest $L$ rays, where $(\phi_\ell^A, \theta_\ell^A)$ are the azimuth and elevation angles of arrival, while $(\phi_\ell^D, \theta_\ell^D)$ are the azimuth and elevation angles of departure. And $a_\ell$ is the path gain for the $\ell$-th ray, and $\tau_\ell$ is the time of arrival. We deploy $4\times2$ ($N_\mathrm t = N_\mathrm r = 8$) uniform planar arrays at both the transmitter and receiver sides. We define $g(\cdot)$ as the pulse shaping filter, and approximate the channel matrix $\mathbf{H}[n]$, $n = 0, 1, \cdots, L_c - 1$ by the geometric channel model
\begin{align}
\mathbf{H}[n]= \sqrt{N_\mathrm t N_\mathrm r}\sum_{\ell = 1}^L g(nT - \tau_\ell) \mathbf{a}_\mathrm r(\phi_\ell^A, \theta_\ell^A)\mathbf{a}^*_\mathrm t(\phi_\ell^D, \theta_\ell^D) a_\ell,
\end{align}
where $T$ is the symbol period, $\mathbf{a}_\mathrm r(\phi_\ell^A, \theta_\ell^A)$ and $\mathbf{a}^*_\mathrm t(\phi_\ell^D, \theta_\ell^D)$ are the steering vectors at the arrival and departure sides for uniform planar arrays.
We apply DFT codebook for the precoder and combiner \cite{YanYanHan:DFT-based-beamforming-weight-vector:10}. Therefore, there are in total of $N_\mathrm B = N_\mathrm tN_\mathrm r= 64$ different beam pairs in our dataset. Assume the $i$-th beam pair, $(\mathbf{w}_i,  \mathbf{f}_i), i\in \{1, 2, \cdots, N_\mathrm B \}$, is selected from the codebook, the received power $y_i$ can be calculaed as 
\begin{align}\label{equ:rxpower}
y_{i}= \sum_{n = 0} ^{L_c - 1} \big |\mathbf{w}_i^*\mathbf{H}[n]\mathbf{f}_i\big|^2. 
\end{align}
\iffalse 
After \emph{raveling} the $N_tN_r \times N_t N_r$ received power matrix $\mathbf{Y}(i,j) = y_{i,j}$, we formulate a vector $\mathbf{y}$, where $\mathbf{y}_{N_tN_r(i-1) + j} = \mathbf{Y}(i,j)$ of length $N = (N_tN_r)^2$. Hence, the optimal beam pair index is $s = \mathrm{argmax}_{i\in\{1, \cdots, N\}}\mathbf{y}_i$.
We assume a least-square approach to estimate the channel as in \cite{VaChoShi:Inverse-Multipath-Fingerprinting:17} (the details are referred to \cite{VaChoShi:Inverse-Multipath-Fingerprinting:17}). The received power $y_i$ for the $i$-th beam pair is calculated as the power of the estimated channel $\hat{\mathbf{h}}_i$, i.e., $|\hat{\mathbf{h}}_i|^2$.\fi
The training label for beam power regression is $\mathbf{y} = [y_1, y_2, \cdots, y_{N_\mathrm B}]$, and the corresponding optimal beam pair can be derived by $s = \mathrm{argmax}_{i\in\{1, \cdots, N_\mathrm B\}}y_i$. 
\subsection{Power quantization} \label{sec:quantization}
In mmWave systems, after beam sweeping is implemented, the infrastructure cannot obtain the \emph{exact} value of received power by channel feedback. Generally, only the quantized CQIs along with the corresponding beam pair indexes are fed back to the infrastructure. Hence, the continuous received power from simulation, which is obtained from (\ref{equ:rxpower}), in \emph{first} row of Table \ref{table:dataset} and represented by $\ell_i$, needs to be quantized by some certain quantization rule. There is a vast body of literature discussing different mapping schemes from received power to CQI \cite{Hua:Interference-measurement-resource:12}, \cite{LeeHanZha:MIMO-technologies-in-3GPP:09}. In Long Term Evolution (LTE), CQI is an indication of what modulation and coding scheme (MCS)/transport block the UE can reliably receive. Specifically, the UE determines the highest MCS for which the block error rate is under 10$\%$, in the bandwidth in which the CSI reference signal is received. Our case is different from the CQI calculation in LTE, since we are not primarily concerned about the appropriate selection of MCS at this point. Instead, we target at predicting the beam power, and transmitting the precise information of the received power to the infrastructure for regression. Therefore, in our case, CQI is a direct indicator of reference signal received power (RSRP). For simplicity, we assume a simple \emph{uniform quantization scheme}, where we use $P_u$ and  $P_\ell$ to \emph{upper} and \emph{lower bound} the power. We then define the CQI \emph{granularity} as $r_\text{CQI}$, and the relationship between the received power $p$ and the CQI index $q$ is defined by 
\begin{align}\label{equ:adc}
\mathcal{Q}(p) =\ceil*{\min\left\{\max\left\{\frac{p - P_\ell}{r_\text{CQI}},  0\right\}, \frac{P_u - P_\ell}{r_\text{CQI}}\right\} }, 
\end{align}
where the idea is to upper bound the received power by $P_u$ (CQI = $\mathcal{Q}(P_u)$) and lower bound it by $P_\ell$ (CQI = 0), and then quantize evenly for the power lying in the range $p\in[P_\ell, P_u]$. And correspondingly, we recover the continuous received power from the CQI by
\begin{align}\label{equ:dac}
r(q) = r_\text{CQI}q  + P_\ell.
\end{align}
After CQI quantization, the entropy of the information is reduced, and quantization inaccuracy is introduced, especially for the power out of the range $[P_\ell, P_u]$, which is either \emph{upper} or \emph{lower} bounded.  In Section \ref{sec:numerical}, we will show that the learning accuracy depends on the aforementioned parameter $P_u$, $P_\ell$ and the quantization granularity. Even with low quantization resolution, however, we show that performance is not degraded significantly.
\begin{table}
	\centering
	\caption{An example of the dataset and the post-processing}\label{table:dataset}
	\begin{tabular}{|c|c|}
		\hline
		\hline
		Original & $\ell_i = y_1,  \cdots, y_{N_\mathrm B}$\\\hline
	CQI & $\ell_c = \mathcal{Q}(y_1), \cdots, \mathcal{Q}(y_{N_\mathrm B})$\\\hline
		Regressor& $\ell_r = r(\mathcal{Q}(y_1)),  \cdots, r(\mathcal{Q}(y_{N_\mathrm B}))$ \\
		\hline
		Ordered beam & $\ell_m(N) = r(\mathcal{Q}(\bar{y}_1)), \cdots, r(\mathcal{Q}(\bar{y}_M))$\\
		\hline\hline
	\end{tabular}\label{table}	
	\vspace{-0.2in}
\end{table}

\iffalse
\begin{figure}
	\includegraphics[width = 3.0in]{simulation_setting.png}
	\caption{The illustration of simulation setting in Wireless Insite. There are in total of two types of vehicles: trucks and cars. We assume the transceivers are mounted on top of the cars only (since trucks are free of blockage and always have LOS links). The buildings are generated of random shapes on both of the two road sides. The green box is the RSU, which we assume is attached to the street lamp, at a height of 5m. }
\end{figure}
\fi

\section{Learning model}
In this section, we explain the rule of encoding the situational features and the approach to predict the beam power. We show that the power prediction is able to predict \emph{any} beam, e.g.,  the strongest beam, second strongest beam, etc. It is also applicable to the predict the beam power based on the beam pair index. 
\iffalse 
\subsection{Deep neural networks} We use the deep learning model based on Keras \cite{keras_web}. We select \emph{mean squared error (MSE)} as the loss function, which is defined as
\begin{align}
\ell_{\text{MSE}}(\mathbf{y}, \hat{\mathbf{y}})  = \frac{1}{|\mathbf{y}|} \sum_{i= 0}^{|\mathbf{y}| - 1} (\mathbf{y}_i - \hat{\mathbf{y}}_i)^2, 
\end{align}
where $\mathbf{y}_i, \hat{\mathbf{y}}_i$ are the real and predicted label, respectively. 
A stochastic gradient descent optimizing algorithm - \emph{Adam} \cite{adamopt} is used to learn the parameters from the model. 

\fi
\subsection{Encoding the geometry}\label{sec:feature}
\begin{figure}
	\includegraphics[width = 3.5in]{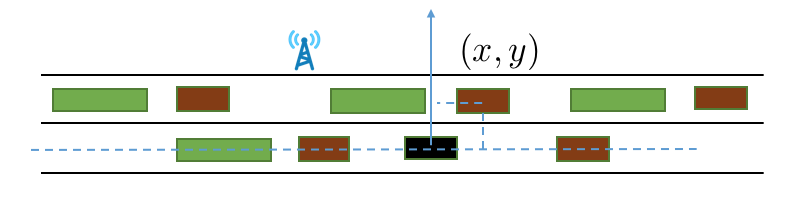}
	\caption{A demonstration of the vehicle's geometry and encoding. The black car is the receiver. The brown boxes are the small-size cars surrounding the receiver. The green boxes denote the surrounding trucks. \label{fig1:encoding}}
\end{figure}

There are different ways to encode the vehicles' situational awareness. Since we deploy two types of vehicles (cars and trucks) randomly on a two-lane street, we need to design the appropriate scheme to encode the geometry and order the features accordingly. Specifically, we apply simple \emph{Cartesian coordinate} to encode the locations, as demonstrated in Fig. \ref{fig1:encoding}. The origin of the Cartesian coordinate is set as the receiver, and the road-side unit and the surrounding vehicles are encoded accordingly under the current coordinate. Also, we can observe from the vehicle deployments that generally for a receiver, it is easier to be blocked and effected by the vehicle on the first lane, i.e., the lane closer to the RSU. Furthermore, large trucks and vehicles that are closer by have more impact on the receiver's beam.  Hence, we propose the following strategy to encode and order the geometry. 

The feature $\mathbf{v}$ is a one-dimensional vector and can be generated by 
\begin{align}\label{equ:feature}
\mathbf{v} = [\mathbf{r}, \mathbf{t}_1, \mathbf{t}_2, \mathbf{c}_1, \mathbf{c}_2].
\end{align}
In (\ref{equ:feature}), $\mathbf{r}$ is the location of the RSU in the Cartesian coordinate, $\mathbf{t}$ represents the truck and $\mathbf{c}$ denotes the low-height cars. The subscripts $1$ and $2$ denote the lane index where the vehicle is located on. For the trucks on the first lane, e.g., given the coordinates of the trucks along the $x$-axis on the first lane in the Cartesian coordinate as $[x_1, x_2, \cdots, x_n]$,   $\mathbf{t}_1$ is composed of the locations of the trucks in the following order. 
\begin{align}
\mathbf{t}_1 &= [x_{i_1}, y_{i_1}, x_{i_2}, y_{i_2}, \cdots, x_{i_N}, y_{i_N}], \nonumber\\
& |x_{i_1}|<|x_{i2}|<\cdots<|x_{iN}|. 
\end{align}
Here, we constrain the number of trucks/cars on each lane as the maximum number of $N$ in order to make dimensions of features consistent under different deployment scenarios. If the number of trucks $n>N$, we \emph{delete} the locations of $N-n$ trucks that are located far away from the feature; otherwise, we add in $n-N$ \emph{virtual} trucks that are lying very far away, where $x = \infty$ (we choose $x = 10^4$ here).  Similarly, the trucks/cars on the different lanes can be encoded.

\subsection{Practical issues with feedback}\label{sec:partialfeedback}
In implementations, the feedback link conveys the information of only a subset of the beam pairs. Generally, the feedback includes the best $M$ beams' received power and the corresponding beam pair index. With limited information of the beam pairs' power, we rearrange the beam pairs in decreasing order of their powers, i.e.,  $\bar{y}_1>\bar{y_2}, \cdots, >\bar{y}_M \cdots, >\bar{y}_{N_\mathrm B}$, and only apply regressor to the first $M$ beams' received power, as shown in the \emph{fourth} row of Table \ref{table:dataset}. The model eliminates the necessity of feeding back information of all beams and can be combined with other models that can rank the beams correspondingly, to achieve even lower overheads. In this paper, we focus on the analysis of only the beam pair with the \emph{strongest power}, i.e., $M = 1$. 

 We also consider the case that \emph{unordered} beam power, as shown in the \emph{first} row of Table \ref{table:dataset}, needs to be fed back to the infrastructures. Larger overheads are introduced to the system and a longer time is required to finish the database establishment. The \emph{full knowledge} of the beams' power, however, provides an easy way to select and recommend the optimal beam pair, and also to evaluate the system performance.  
\section{Performance evaluation}\label{sec:numerical}
In this section, we train the regressors  with different learning models and datasets, to predict the power of each beam. We define the relevant performance metrics, and evaluate the system performance with different features and CQI quantization parameters. Then we examine the performance when power is predicted per beam pair. 
\subsection{Performance metric definition}\label{equ:metric}
Let $\max\{\cdot\}$ denote the maximum element of the vector $(\cdot)$, and $\arg\max\{\cdot\}$ the index of the maximum value and $\mathbbm{1}(\cdot)$ is the indicator function. Given the real power $\mathbf{Y}= \{\mathbf{y}_1, \cdots, \mathbf{y}_m\}$, and the predicted power $\hat{\mathbf{Y}} = \{\hat{\mathbf{y}}_1, \cdots, \hat{\mathbf{y}}_m\}$, the \emph{alignment probability} can be formulated as $P_\mathrm A = \frac{1}{m}\sum_{i=1}^{m}\mathbbm{1}(\arg\max \{\mathbf{y}_i\} = \arg\max\{\hat{\mathbf{y}}_i\})$. 
And we can further define \emph{achieved throughput ratio} $R_\mathrm T$ as
\begin{align}
R_\mathrm T =\frac{\sum_{i = 1}^m  \log_2 ( 1 + \mathbf{y}_i[\arg\max\{\hat{\mathbf y}_i\}])}{\sum_{i=1}^m\log_2(1  + \max\{\mathbf{y}_i\})}.
\end{align}
Achieved throughput ratio $R_\mathrm T$ indicates the system performance in throughput when the system is deployed simply relying on the learning model without beam training. 
\subsection{Regression models}\label{sec:regressionmodel}Using the situational features defined in Section \ref{sec:feature}, we compare results with different regression models. We utilize the root mean squared error (RMSE) to quantify the regression accuracy over the \emph{strongest beam power}, i.e., $\bar{y}_1$, in dB scale.
\iffalse  which is defined as 
\begin{align}
\mathrm{RMSE} = \sqrt{\frac{\sum_{i=1}^n (p_i - a_i)^2}{n}}, 
\end{align}
where $p_i$ and $a_i$ are the predicted power and the true power respectively, and $n$ is the number of the test samples.\fi 
\begin{table}[]
	\centering
	\caption{Regression RMSE over the strongest beam power using different regression algorithms.}
	\label{table:rmse}
	\begin{tabular}{|l|l|}
		\hline
		\centering
		& RMSE (dBm)\\\hline
		\centering
		Linear regr&6.199  \\\hline
		SVR & 3.645   \\\hline
		Random Forest &  1.726 \\ \hline
		Gradient Boosting & 2.814   \\\hline
	\end{tabular}
\end{table}
Specifically, we compare the prediction results among linear regression, support vector regression, Random Forest regression and gradient boosting regression in Table \ref{table:rmse}. It is shown that the Random Forest is a good fit for our specific dataset, since it is able to implicitly select the features and generalizes well by ensembles. Also, the Random Forest is fast to train and is a promising learning method that could be applicable in industry field implementations. 

\subsection{Different levels of situational awareness}
In this section, we show how situational awareness can help to predict the beam power. Current pathloss model relies on the relative distance \cite{RapSunMay:Millimeter-wave-mobile:13} or the absolute locations of the receiver and the transmitter \cite{WanVenHea:MmWave-vehicle-to-infrastructure-communication::18}. Result in \cite{VaChoShi:Inverse-Multipath-Fingerprinting:17} also showed that the receiver location only can provide useful information about the beam by exploiting dataset of previous transmissions. Here we show that in an urban vehicular context, the environment information of vehicle locations could be leveraged for more accurate prediction of the beam power. 
\begin{figure}
	\centering
	\includegraphics[width = 2.6in]{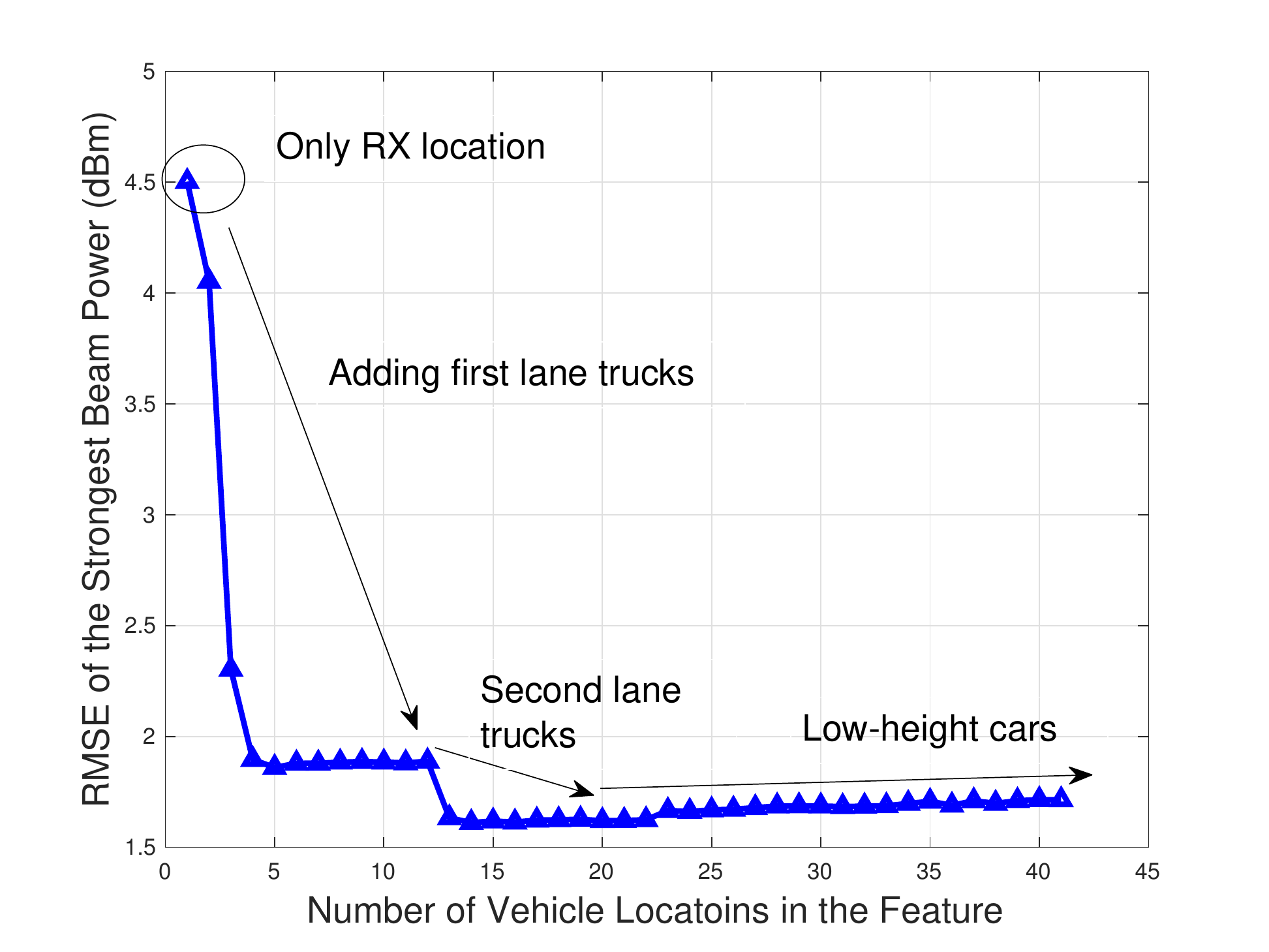}
	\caption{Comparison of alignment with different levels of situational awareness. }\label{fig:sa}
\end{figure}
 Based on the vehicle location \emph{order} in Section \ref{sec:feature}, in Fig. \ref{fig:sa}, we plot the RMSE of the strongest beam pair using the Random Forest regressor with the \emph{first} $i$-th vehicles' locations in the feature as explained in Section \ref{sec:feature}. It could be observed that the first lane trucks' locations provide abundant information about the beam power. RMSE is reduced from 4.5 to around 1.7, compared to the case when only the receiver location is used as the feature. More truck locations on the second lane finally reduce the RSME to around 1.6. Low-height cars, however, do not further help beam prediction. Slight degradations of RMSE are shown when extra cars' locations are fed in the feature. Hence, we conclude that the high trucks' locations, especially those on the first lane, are more relevant in predicting the beam power and a concise set of location features are sufficient in providing good predictions. 

\subsection{CQI Quantization}\label{sec:result_quant}
\begin{figure*}
	\centering
	\begin{minipage}{0.45\textwidth}
		\centering
		\includegraphics[width = 2.6in]{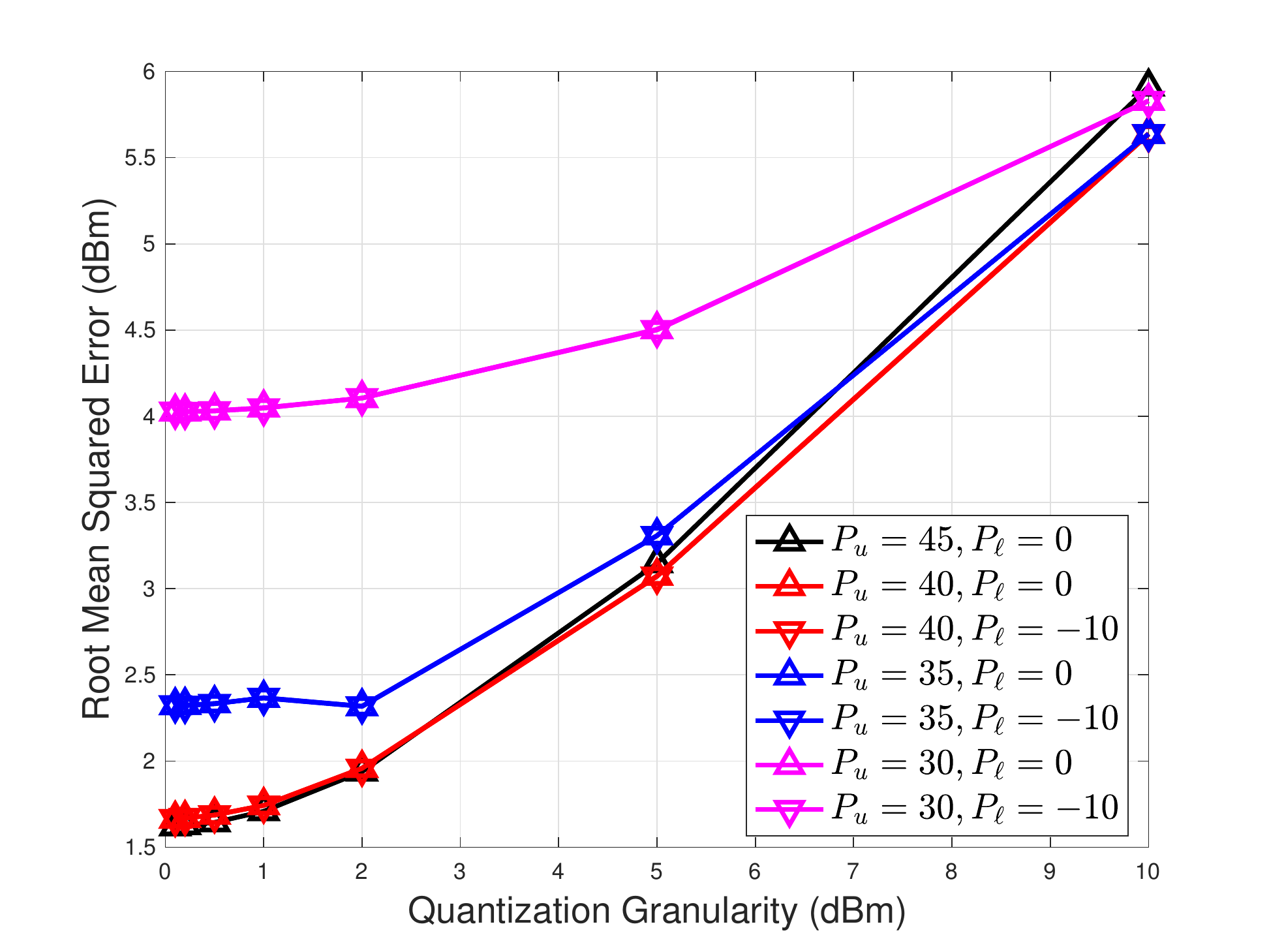}
		\caption{Comparion of RMSE using different selections of quantization parameters $P_u$, $P_\ell$ and $r_\text{CQI}$. Specifically, we consider some combinations of $P_u = 45, 40, 35, 30$ dBm, and $P_\ell = 0, -10$ dBm and quantization granularity $r_\text{CQI} = 0.1, 0.2, 0.5, 1, 2, 5, 10$ dBm. }\label{fig:quantization}
	\end{minipage}	
	\hspace{0.5cm}
	\begin{minipage}{0.45\textwidth}
		\centering
		\includegraphics[width = 2.6in]{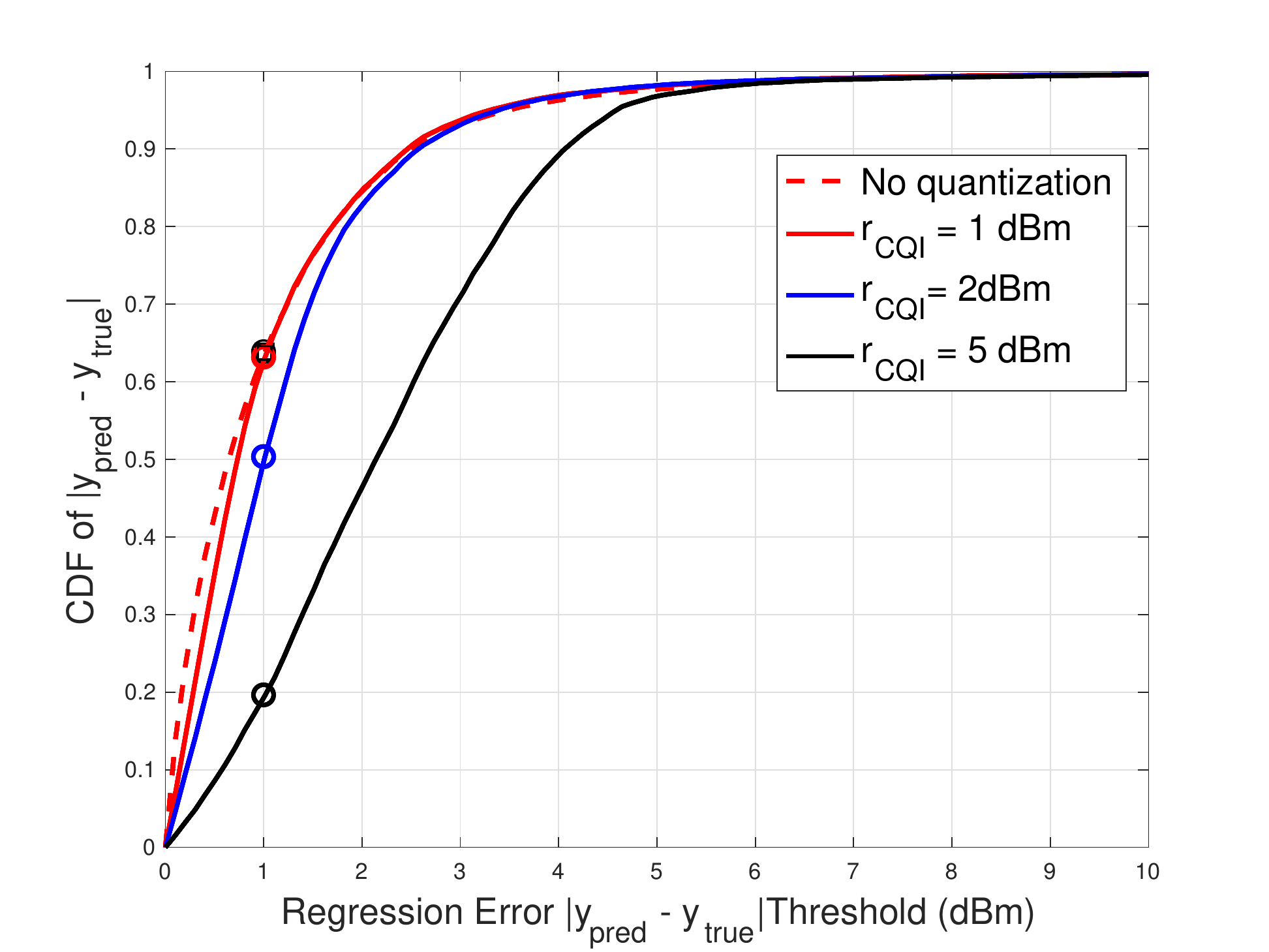}
		\centering
		\caption{Comparison of regression error $|y_\text{pred} - y_\text{true}|$ with different CQI quantization granularities $r_\text{CQI} = 1, 2, 5$ dBm and the case without CQI quantization. The $y$ coordinates of the three circles (in red, blue and black) represent the probability that the regression error is small than $1$ dB. }\label{fig:cdf}
	\end{minipage}
\end{figure*}

In this section, we compare the performance of the strongest beam pair power prediction with CQI quantization. Specifically, we quantize the power first and recover the continuous power as explained in Section \ref{sec:quantization}. We evaluate the RMSE of prediction using different combinations of parameters $r_\text{CQI}$, $P_u$ and $P_\ell$. It is observed from the dataset that the highest beam power across the dataset is $P_{\max} = 44.23$ dBm and the lowest power is $P_{\min} = -15.15$ dBm. Based on these, we select the combinations of $P_u$ and $P_\ell$ as shown in Fig. \ref{fig:quantization}. It is shown that the RMSE increases with a larger granularity generally. In the small quantization granularity regime, e.g., $r_\text{CQI}  = 0, 0.1, 0.2, 0.5, 1$ dBm, there are no significant differences among the RMSEs. Also, the larger upper bound $P_u$ gives more accurate predictions, while the lower bound $P_\ell$ has negligible impact. The reason is that the power of the strongest beam power is generally large and a \emph{small upper bound} will bring large errors by quantization. It should be noted, however, that both the upper and lower bound needs to be carefully designed in order to guarantee the quantization accuracy with the given statistics of the datasets. 

Fig. \ref{fig:cdf} further compares the cumulative density function (CDF) of the regression error $|y_\text{pred} - y_\text{true}|$ with different quantization granularities. The three dots in red, blue and black indicate the probabilities that the regression error is smaller than $1$ dBm. It is shown that with $r_\text{CQI} = 1$ or $2$ dBm, we can still guarantee error of more than $50\%$ of the predictions to be smaller than $1$ dBm. And there is barely any difference between the accuracy using $1$ dBm quantization and the case without quantization. 

\subsection{Regression over all beam pairs}
\begin{table}[t]
	\centering
	\caption{Alignment probability and achieved throughput ratio with different CQI quantization granularities.}
	\label{my-label}
	\begin{tabular}{|l|l|l|}
		\hline
		\hline
		\centering
		& $P_\mathrm A (\%$)& $R_\mathrm T (\%)$ \\\hline
		Classifier & 84.6 &  98.4\\\hline
		No QT& 82.0& 98.8 \\\hline
		$r_\text{CQI} = 1$ & 81.1 &  98.8\\\hline
		$r_\text{CQI} = 2$ &  79.2& 98.7\\ \hline
		$r_\text{CQI} = 5$ & 72.6 & 97.8 \\\hline
	\end{tabular}
\end{table}
Previous results apply a regression model over the strongest beam power. In this section, we predict the power based on the beam pair index as defined in (\ref{equ:rxpower}). The advantage of this method is that it gives extra information to the order of the beam pair power, and helps to select the beam pair for data transmission. This case, however, requires the knowledge of the power of \emph{all} beam pairs, which introduces larger overheads to establish the dataset. Here, we focus on evaluating how the power prediction can be helpful in assisting the beam selection, without considering the overhead issues. We compare the alignment probability $P_\mathrm A$ and the achieved throughput $R_\mathrm T$ as defined in Section \ref{equ:metric}, using different quantization granularities of CQI and the case when we simply apply a classifier over the optimal beam pair index. Similarly as Section \ref{sec:result_quant}, quantization with high resolution does not introduce a lot of performance degradation. Also, even though the alignment probabilities $P_\mathrm A$ with the regression models are lower than the classifier, the achieved throughputs are higher than that of the classification. More beams' power provides intrinsic information about the beam power orders. Lastly, the achieved throughput ratios are all very high due to the fact that there are not big differences among the power of the top beams. The results show that our model is good at identifying the ``good" beams from the ``bad" beams, even though 100$\%$ alignment probability cannot be achieved.

\section{Acknowledgments}
This research was partially supported by a gift from Huawei through UT Situation-Aware Vehicular Engineering Systems (UT-SAVES), and by the U.S. Department
of Transportation through the Data-Supported Transportation
Operations and Planning (D-STOP) Tier 1 University
Transportation Center and
Communications and Radar-Supported Transportation Operations
and Planning (CAR-STOP) project funded by the Texas
Department of Transportation. 
\iffalse
Fig. \ref{fig:errorcdf}
\begin{figure}
	\centering
	\includegraphics[width = 2.8in]{figure_compare_classification.eps}
	\caption{Comparison of coverage probability of using regressor with different resolutions of CQI and classifier based on random forests.}
\end{figure}
\fi

\iffalse 
2. compare with the results of different learning algorithms  --- probably the throughput or simply the MSE of different predictions of different beam s

3. Comparison of utilizing random forest when we are giving different CQI for the transmission 

4. Combining the results of power prediction and beam prediction to formulate a joint opitmizaiotn problem possibly with a simple objective function. For example, to optimize the joint throughput. 

5. compare with different antenna size and the beam codebook size

\emph{Overhead} We assume exhaustive beam sweeping is implemented for the data collection. Therefore, the majority of the overhead comes from the feedback of CQIs. We evaluate the feedback overhead by \emph{bits/user}. For a 

4.

\newpage
\fi
\footnotesize
	\bibliographystyle{ieeetr}
\bibliography{refer}

\end{document}